\documentclass{emulateapj}

\slugcomment{}

\shorttitle{Chemical anomalies and extreme blue HB stars}
\shortauthors{Carretta et al.}

\begin{document}


\title{The link between chemical anomalies along the red giant branch and the 
horizontal branch extension in globular clusters}


\author{Eugenio Carretta\altaffilmark{1}, Alejandra Recio-Blanco\altaffilmark{2}, 
Raffaele G. Gratton\altaffilmark{3},
Giampaolo Piotto\altaffilmark{4}, Angela Bragaglia\altaffilmark{1} }  

\altaffiltext{1}{INAF-Osservatorio Astronomico di Bologna, via Ranzani 1, 
I-40127 Bologna, Italy. email: eugenio.carretta@oabo.inaf.it,angela.bragaglia@oabo.inaf.it}
\altaffiltext{2}{Dpt. Cassiop\'ee, Observatoire de la Cote d'Azur, CNRS/UMR 6202, BP 4229, 
06304 Nice Cedex 4, France. email: Alejandra.RECIO-BLANCO@obs-nice.fr}
\altaffiltext{3}{INAF-Osservatorio di Padova, Vicolo dell'Osservatorio 5, 
I-35122 Padova,Italy. email: raffaele.gratton@oapd.inaf.it}
\altaffiltext{4}{Dipartimento di Astronomia, Universit\`a di Padova, Vicolo dell'Osservatorio 2, 
I-35122 Padova,Italy. email: giampaolo.piotto@unipd.it}


\begin{abstract}

We find a strong correlation between the extension of the
Na-O anticorrelation observed in  red giant branch (RGB) stars and the high
temperature extension of the  horizontal branch (HB)  blue tails of Galactic
globular clusters (GCs).  The longer is the O-depleted tail of the Na-O
anticorrelation  observed in the  RGB stars, the higher is the maximum
temperature reached by  the bluest HB stars in the GC.  This result provides a
clear, empirical evidence of a link between the extension of the HB and the
presence of  star-to-star abundance variations of proton-capture elements in GC
stars. We discuss the possible interpretation of this correlation.

\end{abstract}


\section{Introduction}
Since more than thirty years, we know that globular cluster (GC) stars show a
dispersion in the content of light elements like C, N, O, Na, Mg, Al. The
lightest C and N are anticorrelated both in field and cluster stars (Smith  \&
Martell 2003). The environment of GCs must be
responsible for the bulk of chemical anomalies observed among stars in GCs (at
odds with field stars, Gratton et al. 2000), where the heavier proton-capture
elements are anticorrelated (Na vs O and Mg vs Al, respectively; see Gratton,
Sneden, \& Carretta 2004  for extensive references).

The theoretical background (e.g. Denisenkov \& Denisenkova 1990, Langer et al.
1993) points to the simultaneous run of the CNO, NeNa and MgAl cycles of
H-burning at high temperature. In turn, this  implies that the observations of
the same anticorrelations in unevolved cluster stars (Gratton et al. 2001;
Ramirez \& Cohen 2002; Carretta et al. 2004, 2005; Cohen \& Melendez 2005) can
be explained only if the involved proton-reactions occurred in more massive
stars  of a previous generation than the presently observed one, and the
anomalies concern a large fraction of the stellar structure, non simply a
surface layer (Cohen et al. 2002). Multiple star formation episodes in GCs have
eventually been directly observed in the form of a main sequence split in
$\omega$ Centauri (Bedin et al. 2004) and NGC 2808 (Piotto et al. 2007), and in
a subgiant branch split in NGC 1851 (Milone et al. 2007).

Red giant branch (RGB) stars with both normal and anomalous composition must
eventually end up on the Zero Age Horizontal Branch (ZAHB) after the onset of
core He-burning, and we might expect to track their different chemical signatures
in the ensueing evolution.  On one hand, the  horizontal branch (HB) phase is an
ideal locus to investigate the effects of chemical anomalies on the stellar
structure, as it acts as an amplifier of the  physical conditions consequent to
the star's previous evolution. On the other hand, the location of stars onto the
ZAHB is sensitive to a number of secondary parameters (age, CNO composition,
helium content, rotation etc, see e.g., Fusi Pecci et al. 1993) beside the main
parameter, metallicity,  to make it difficult to disentangle the contribution of
the single one.  A combination of these parameters actually conspire to give a 
variety of HB morphologies. 

Recently, Recio-Blanco et al. (2006;  hereafter RB06) provided a quantitative
estimate of the temperature extent of the HB by using the homogeneous set of 
color-magnitude diagrams (CMDs) of the HST snapshot program by Piotto et al.
(2002). RB06 found a  significant correlation between the HB extension and  the
cluster total mass. They interpreted this correlation as an observational
evidence of self-pollution, assuming, as suggested by D'Antona et al. (2002),
that the factor responsible for the HB extension might be an He enhancement predicted
to accompany  the observed CNONaMgAl star-to star abundance variations on the
RGB.

In this Letter we present a tight and direct empirical correlation between the
maximum temperature along the ZAHB and the extent of the Na-O and Mg-Al
anticorrelations, as defined by Carretta (2006), and discuss its implications.

\section{The data sample}

The parameter $\log T_{\rm eff}$(HB) was defined in RB06 as the maximum
temperature reached by HB stars in each cluster. Briefly, ZAHB models by Cassisi
et al. (1999) were fitted to  the CMDs of 54 GCs  from the $HST$ snapshot survey
by Piotto et al. (2002). The RB06 parameter well compares with similar
quantities, such as the HB length ($L_t$, see Fusi Pecci et al. 1993), but it
has the advantage of having been extracted from a  photometrically homogeneous
catalog of CMDs.
Table~1 lists the $\log T_{\rm eff}$(HB) values for the subset of clusters 
discussed in the present paper.

To give a quantitative definition of the extension of the anticorrelations
observed among RGB stars, Carretta (2006) proposed to use the interquartile
range (IQR) of the [O/Na]  and [Mg/Al] abundance ratios. This quantity
(see Cleveland 1993 for a description of the technique)
is especially useful for indicating
whether a distribution is skewed and whether there are potential outliers. 

The IQR values for the [O/Na] and, when available, [Mg/Al] ratios come  from
three sources. Those for NGC~362, NGC~5904 (M~5), NGC~6205 (M~13), NGC~6838
(M~71),  and NGC~7078 (M~15) are taken from Carretta (2006), who used the
largest samples available in literature from high-resolution spectroscopic
analysis.  The original data sources are given therein and not repeated here.
A second set of values is derived (only for IQR[O/Na]) from the very large
samples recently analyzed by our group using 
FLAMES-GIRAFFE spectra. These are available for NGC~2808 (90 stars, Carretta 
et al. 2006),
NGC~6441 (25 stars, Gratton et al. 2007), and NGC~6218 (M~12, 90 stars,
Carretta et al. 2007a).
To this GIRAFFE subsample we add NGC~6388 (Carretta et al. in preparation), 
with about 30 stars observed. Since 
for that program and instrumental set up (see Carretta et al. 2006 for
details) only O and Na abundances are available, we adopt for NGC~2808
the value of IQR[Mg/Al] based on the 19 FLAMES-UVES spectra
analyzed by Carretta (2006).
The third set of values for IQRs of both [O/Na] and [Mg/Al] is taken from the
analysis (Carretta et al. in preparation) of FLAMES-UVES spectra taken
simultaneously with the GIRAFFE spectra:  for the present paper we computed the
IQRs for NGC~104 (47 Tuc), NGC~1904 (M~79), NGC~3201, NGC~4590 (M~68), 
NGC~6397, NGC~7099 (M~30). The IQR for [Mg/Al] in NGC~6218 is also derived from
this analysis. In this case, numbers are more limited,  with a
maximum of 14 stars in  each cluster.

Appropriate Monte-Carlo simulations show that uncertainty estimates for the IQRs
due the stochastic errors is given by the ratio between the inter-quartile value
divided by the square root of the number of stars used to derive the relation,
times a factor quite close to 1 which depends on the shape of the distribution
(e.g. it is 1 for a uniform distribution and 1.17 for a Gaussian distribution).
For [O/Na], the typical uncertainties in IQR's range from 0.07 dex (NGC6397), to
0.33 dex (NGC1904), with a median value of 0.15 dex. Similar values are obtained
for [Mg/Al].

Values of IQRs of [O/Na] and [Mg/Al] ratios,  the adopted metallicity [Fe/H],
and the source of data are listed in Table 1, together with total visual
luminosity $M_V$ and HB ratio parameter HBR=(B-V)/(B+V+R) taken from the updated
version of the Harris (1996) catalog. The normalized age parameter comes from
the compilation of De Angeli et al. (2005).

\section{A strong link between the Na-O anticorrelation and the maximum
temperature in HB}

In Fig.~1, the maximum temperature reached along the HB in each cluster is
plotted as a function of the IQR parameter for the NaO and MgAl distributions.
Interestingly enough, the resulting correlations are very tight, comparable to 
the bivariate correlations found with metallicity and total luminosity by RB06
(see e.g., their Fig.~8). The Spearman rank correlation coefficient is
$r_s=0.828$ for the correlation with IQR[O/Na] (15 clusters) and $r_s=0.831$ with
IQR[Mg/Al] (12 clusters). The one-tailed test gives a negligible probability
that this is a mere chance result (e.g., $7 \times 10^{-5}$ for [O/Na]).

Fig.~1 provides: (i) a clear empirical evidence of a link between the amount
of abundance variations and the length of the blue HBs, with clusters with more
pronounced Na-O (and simultaneous Mg-Al) anticorrelations having HBs reaching
higher temperatures in their blue tails and (ii) it tells us that the IQR value
contains in itself as much physical information as the combined effect of total
mass and metallicity (see RB06). How can these unequivocal observational facts
be explained?

RB06 found that more massive clusters tend to show HBs more extended to higher
temperature. This is well understood as a better capability of deeper potential
wells to retain ejecta from stars in a self-pollution scenario, which is the
currently most popular way  to explain the observed cluster abundance variations
(see, e.g., D'Antona et al. 2002, Gratton et al. 2004, Carretta et al. 2005). He
enhancement is expected to be present in the stars born out from the gas
composed by a mixture of un-processed  and polluted matter  coming from a
previous generation of stars. The most favourite classes of candidate polluters
(namely, fast rotating massive main sequence stars, Decressin et al. 2007, and
intermediate-mass AGB stars, D'Antona et al. 2002)  can both potentially pollute
the existing interstellar material with products of complete CNO burning,
including He. In this framework, D'Antona et al. (2002) invoked a spread of He
as the key ingredient to naturally reproduce the whole HB morphology in GCs.

The implied bottom line is that  there must be a one-to-one correspondence
between the chemical composition of stars on the RGB and their successive
location on the HB, in the sense that He enhancement should also provide the
needed  difference in the mass of He-burning stars required to spread
them from the red HB  (stars with normal composition) to the hottest part of the
HB (stars with extreme abundance alterations). The correlation of Fig.~1 seems
to confirm this idea, but  also  highlights other important pieces of 
information which needs further investigation, and which will help to better
understand the effects of chemical anomalies resulting from the pollution
scenario.

We start by noting that the linear correlation in Fig.~1 is even more
significant than the monovariate relation between the HB extension and the total
mass found by RB06. This is not entirely unexpected, since total mass should be
used only as a good $proxy$ of the effect of self-enrichment in GCs, because the
dynamical evolution of a GC is also related to its interaction with the Galactic
potential. Not surprisingly, Carretta (2006) showed that the best correlations
involving the IQR values were not with the total mass, but with the cluster
orbital parameters. This can be well explained by comparing the typical orbital
periods in the Galaxy and the typical timescales involved in the  release of
yields by candidate polluters: both are of the order of a few $10^7-10^8$
yrs,  depending on the mechanism responsible for the pollution. It
follows that GCs with eccentric orbits may spend long periods of time relatively
undisturbed by bulge and/or disk shocks, and by interactions with the giant
molecular clouds in the Galactic disk. Therefore, they have enough time and can
efficiently use their whole gas content to build up a second stellar
generation.  On the other hand, GCs in less eccentric orbits and with shorter
periods are more affected by passages through the Galactic disk or bulge, with
consequent shocks and gas removal.

The present-day total mass is only a {\it lower limit} for the cluster original
mass, and not necessarely correlated with the gas reservoir at the time when GCs
begun to form a second star generation from polluted matter. Therefore, a simple
mass vs extension of chemical anomalies relation cannot fully account for the GC
evolutionary history, and not necessarely a larger mass should always be
associated to a more pronounced Na-O anticorrelation.

Furthermore, Carretta et al. (2007b) pointed out that a direct  one-to-one
correspondence between the amount of He pollution and HB morphology is not
subtantiated (see their Fig.~5 and related discussion). Hence, the effect of
self pollution must be more complicated, and additional mechanisms must be at
work, as we will outline below.

Here we propose a different interpretation of the strong empirical link found
between the dispersion in light element abundances and the hottest part of the
HB.

Let us consider the case of NGC 2808, where Sandquist \& Martel (2007) recently
found a clear deficit of bright giants on the RGB. A likely explanation is that
extreme mass loss for a fraction of stars near the tip of the RGB may reduce the
envelope mass below a critical threshold, and that these stars may interrupt
their RGB evolution well  below the tip, igniting  core He-burning only at
higher temperature (hot He-flashers, Castellani \& Castellani 1993). As a
consequence, they end up on a very hot ZAHB location, contributing to build up a
conspicuous  blue tail on the HB.  We propose that these stars with extreme mass
loss are also the stars with the extreme values of the Na-O anticorrelations. We
do not observe them  close to the tip of the RGB simply because they already
evolved out of the RGB.

Are there any other  observational facts supporting this interpretation?  The
comparison of M~13 and NGC~2808 is illuminating, in this respect. It is well
known that the most O-depleted giants in M~13 are also among the brightest ones,
all near the RGB-tip (Sneden et al. 2004). On the other hand, in NGC~2808, such
a trend  is not observed. Carretta et al. (2003) found no significant variations
in Na abundances as a function of magnitude in a sample of about 90 stars, from
the bump level up to the RGB-tip in NGC~2808; the same behaviour is seen also
for O abundances  (Carretta 2006). When coupled to the deficit of bright giants
in NGC~2808 noted by Sandquist \& Martel, the inference from this comparison
is that  in M~13 we are probably seeing the  {\it most He-poor among the
Na/He-enhanced stars} that we can actually observe. We propose that He
self-pollution probably reached even higher levels in NGC 2808, so high that
stars with most extreme composition left the RGB well in advance of approaching
the classical tip level. They were able to decrease their envelope masses,
experience an hot He-flash, and  contribute to the  clump of blue, extremely hot
HB stars of this cluster, which is not present at all in M13.

Our proposed scenario might account for cases where the difference between the
ratio of super O-poor stars on the RGB with respect to normal composition stars
is smaller than the ratio of stars on the EHB with respect to global blue HB
population in each cluster (see the case of NGC6752: Carretta et al. 2007b). 
As a caveat, we remind that close binaries evolving through the common envelope
phase should be also taken into account as potential members of the EHB. This
mechanism explains the EHB field stars (the so-called B subdwarfs). However, the
relative weight of this source of EHB stars in GCs should be small, because
direct observations of hot HB stars by Moni Bidin et al.(2007) revealed that the
binary fraction among hot EHB clump is low, at least in NGC~6752. Finally, even
stars of more normal composition making up the classical tip He-flash but with
individual mass loss at the high extreme of the average mass loss range in
clusters ($\Delta$M$\sim0.2$M$_\odot$, with a $\sigma \sim0.02$ M$_\odot$, Lee,
Demarque \& Zinn 1994), might end on the EHB.

Finally, as shown by Carretta (2006), and confirmed by the extended sample of
Table~1, there is no correlation between the Na-O and Mg-Al chemical variations
and the $global$ HBR parameter. In the present paper we show a strong link 
between the extension of the chemical anticorrelation and the extension of the
HB. Apparently, the signature of non-standard chemical composition affects  the
HB extension towards hot temperatures (low envelope masses),  but much less the 
overall distribution of the stars along it. Other parameters must contribute to
the global shape (temperature distribution) of the HB,  and presumably also to
the total extension.

This conclusion is confirmed by the correlations between age and the relevant
parameters in Table 1: we found a rather good correlation of age with the HBR
parameter (correlation -0.51), but only a weak dependence of the HB maximum
temperature on age (correlation -0.09). This strenghtens our proposed scenario,
by defining in a quantitative way that most of the effect on the extreme blue HB
is due to abundance variations and to metallicity. The role of age in shaping
the hottest part of the HB seems to be only a secondary effect, while it seems
to have a  more significant effect on the overall distribution of stars from the
red to the blue ends of the HB.

In summary, we found a very strong correlation linking the pattern of chemical
variations observed for proton-capture elements in globular cluster red giants
and the highest temperature that stars may reach on the ZAHB. A viable mechanism
is enhanced mass loss on the RGB for stars with extreme anomalies in
composition; this may decrease their envelope mass below a critical threshold
and, after onset of core He-burning in a hot flash, may force the star to  spend
the following HB lifetime  at very high temperature. The deficit of bright
giants observed in NGC~2808, as well as the lack of any luminosity dependence
for O-variations in this cluster are naturally accounted for in our
interpretation. We believe that the present correlation is an important step
forward to define  and quantify the link between abundance variations (likely
arising in the first phases of star formation) and the HB morphology. 

\acknowledgments
We acknowledge partial finantial support from PRIN-INAF 2005 
``Experimenting stellar nucleosynthesis in clean environments"

\clearpage

\begin{deluxetable}{rccccrccl}
\tabletypesize{\scriptsize}
\tablecaption{Adopted parameters.\label{parabu}}
\tablehead{\colhead{GC ID}&
\colhead{$\log T_{\rm eff}^{\rm max}$(HB)}&
\colhead{[Fe/H]}&
\colhead{$M_v$}&
\colhead{N.age}&
\colhead{HBR}&
\colhead{IQR[O/Na]}&
\colhead{IQR[Mg/Al]}&
\colhead{source}\\
}
\startdata
NGC 0104 & 3.756 & $-$0.76 & $-$9.42 &  0.97 & $-$0.99 & 0.44 & 0.11 & Carretta et al. (in prep)\\
NGC 0362 & 4.079 & $-$1.16 & $-$8.40 &  0.74 & $-$0.87 & 0.67 & 0.38 & Carretta (2006)\\
NGC 1904 & 4.352 & $-$1.57 & $-$7.86 &  0.90 &   +0.89 & 0.98 & 0.41 & Carretta et al. (in prep)\\
NGC 2808 & 4.568 & $-$1.15 & $-$9.36 &  0.77 & $-$0.49 & 0.99 & 0.98 & Carretta et al. (2006)\\
NGC 3201 & 4.079 & $-$1.58 & $-$7.49 &  0.77 &   +0.08 & 0.82 & 0.45 & Carretta et al. (in prep)\\
NGC 4590 & 4.041 & $-$2.06 & $-$7.35 &  0.92 &   +0.17 & 0.48 & 0.26 & Carretta et al. (in prep)\\
NGC 5904 & 4.176 & $-$1.27 & $-$8.81 &  0.83 &   +0.31 & 0.60 & 0.39 & Carretta (2006)\\
NGC 6205 & 4.505 & $-$1.54 & $-$8.70 &  1.05 &   +0.97 & 0.99 & 1.24 & Carretta (2006)\\
NGC 6218 & 4.217 & $-$1.48 & $-$7.32 &  0.94 &   +0.97 & 0.80 & 0.31 & Carretta et al. (2007a)\\
NGC 6388 & 4.255 & $-$0.60 & $-$9.82 &       &         & 0.80 &      & Carretta et al. (in prep)\\
NGC 6397 & 3.978 & $-$1.95 & $-$6.63 &  1.00 &   +0.98 & 0.23 &      & Carretta et al. (in prep)\\
NGC 6441 & 4.230 & $-$0.53 & $-$9.47 &       &         & 0.68 &      & Gratton et al. (2007)\\
NGC 6838 & 3.763 & $-$0.73 & $-$5.56 &  0.91 & $-$1.00 & 0.35 & 0.16 & Carretta (2006)\\
NGC 7078 & 4.477 & $-$2.26 & $-$9.17 &  0.94 &   +0.67 & 0.79 & 0.80 & Carretta (2006)\\
NGC 7099 & 4.079 & $-$2.12 & $-$7.43 &  1.08 &   +0.89 & 0.27 & 0.50 & Carretta et al. (in prep)\\
\enddata
\tablenotetext{a}{The IQR[O/Na] values in Carretta et al. (in prep) are from UVES spectra, except 
for NGC 6388, from Giraffe spectra.}
\tablenotetext{b}{The IQR[Mg/Al] for NGC 2808 is from Carretta (2006) and for NGC~6218 is
from UVES spectra (Carretta et al. (in prep).}
\end{deluxetable}

\clearpage

\begin{figure*}
\plotone{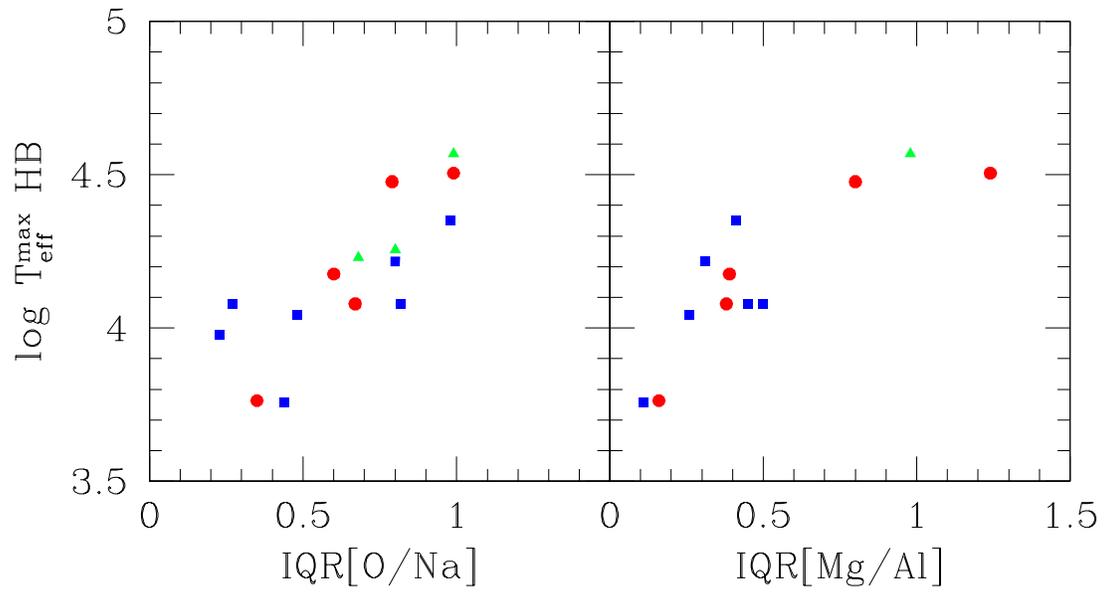}
\caption{Correlations between the maximum temperature reached along the HB and
IQR[O/Na] (left panel) and IQR[Mg/Al] (right panel).
Filled (red)
circles are IQRs from Carretta (2006); filled (green) triangles are IQRs derived
from GIRAFFE spectra and filled (blue) squares from UVES spectra.
\label{anti}}
\end{figure*}

\end{document}